\documentclass{aastex}          
\usepackage{spr-astr-addons}    
\usepackage{epsfig}
\usepackage{graphicx}
\usepackage{color}

\begin{document}
%
\title{Torsional Alfv\'en waves and the period ratio $P_{1}/P_{2}$ in spicules}

\shorttitle{Torsional Alfv\'en waves}
\shortauthors{Ebadi et al.}

\author{H.~Ebadi\altaffilmark{1}}
\affil{Astrophysics Department, Physics Faculty,
University of Tabriz, Tabriz, Iran\\
e-mail: \textcolor{blue}{hosseinebadi@tabrizu.ac.ir}}
\and
\author{S.~Shahmorad}
\affil{Applied Mathematics Department, Mathematics Faculty,
University of Tabriz, Tabriz, Iran\\
e-mail: \textcolor{blue}{shahmorad@tabrizu.ac.ir}}

\altaffiltext{1}{Research Institute for Astronomy and Astrophysics of Maragha,
Maragha 55134-441, Iran.}

\begin{abstract}
The effects of both density stratification and magnetic field expansion on torsional Alfv\'en
waves in solar spicules are studied. Also, their eigenfrequencies, eigenfunctions, and the period
ratio $P_{1}/P_{2}$ are obtained with a novel mathematical method.
We showed that under some circumstances this ratio can approach
its observational value even though it departs from its canonical value of $2$. Moreover,
Eigenfunction height variations show that the oscillations amplitude are increasing towards higher
heights which is in agreement with the observations. This means that with a little increase in height,
amplitude of oscillations expands due to the significant decrease in the density.
\end{abstract}


\section{Introduction}
\label{sec:intro}
Observation of oscillations in solar spicules may be used as an indirect evidence of
energy transport from the photosphere towards the corona.
Transverse motion of spicule axis can be observed by both,
spectroscopic and imaging observations.  The periodic Doppler
shift of spectral lines have been observed from ground based
coronagraphs \citep{nik67,Kukh2006,Tem2007}.
But Doppler shift oscillations with period of $\sim\!\!5$
min also have been observed on the SOlar and Heliospheric
Observatory (\emph{SOHO}) by \citet{xia2005}.
Direct periodic displacement of spicule axes have been
found by imaging observations on Solar Optical Telescope
(SOT) on \emph{Hinode\/} \citep{De2007,Kim2008,he2009}.

The observed transverse oscillations of spicule axes were interpreted by kink
\citep{nik67,Kukh2006,Tem2007,Kim2008,Ebadi2012a} and Alfv\'{e}n \citep{De2007} waves.
The kink mode, amongst others, differs from the torsional Alfv\'{e}n mode,
in that it displaces the whole flux tube in the transverse direction, while torsional
mode does not displace the tube at all. Hence the kink mode is a bulk motion of the internal and external
plasma, whereas the torsional Alfv\'{e}n mode can exist independently on each magnetic surface. However,
despite this significant difference, the kink mode is still highly Alfv\'{e}nic \citep{Verth2010}.
All spicule oscillations events are summarized in a recent review by \citet{Tem2009}.

One of the most important functions of coronal seismology is determining the period ratio $P_{1}/P_{2}$ between the period
$P_{1}$ of the fundamental mode and the period $P_{2}$ of its first harmonic.
\citet{Ebadi2014} analyzed the time series of oxygen line profiles, obtained from SUMER/SOHO on the solar south limb spicules.
They calculated Doppler shifts and consequently Doppler velocities on a coronal hole region.
They performed wavelet analysis to determine the periods of fundamental mode and
its first harmonic mode. The calculated period ratios have departures from its canonical value of $2$.
Different factors such as the effect of density stratification \citep{Andries2009} and magnetic twist \citep{Karami2009,Karami2012}
can cause the deviation of the period ratio from its canonical value.

The waves and the period ratio $P_{1}/P_{2}$ have been studied in coronal loops both theoretically and observationally.
The observed values of this ratio in coronal loops is either smaller or larger than $2$ \citep{Verwichte2004,Andries2009}.
\citet{Srivastava2008} Using simultaneous high spatial and temporal resolution H$\alpha$
observations studied the oscillations in the relative intensity to explore the
possibility of sausage oscillations in the chromospheric cool post-flare loop.
They used the standard wavelet tool, and find $P_{1}/P_{2} \sim 1.68$. They suggested
that the oscillations represent the fundamental and the first harmonics
of the fast-sausage waves in the cool post-flare loop.
\citet{Orza2012} studied the $P_{1}/P_{2}$ period ratio of transversal loop oscillations for the diagnostics of longitudinal structuring of
coronal loops. Their analysis shows that this ratio is sensitive to the temperature difference between the loop and its
environment and this effect should always be taken into account when estimating the degree of density structuring.
\citet{Verth2010} investigated how the radial and longitudinal plasma structuring affects the observational properties of torsional Alfv\'{e}n
waves in magnetic flux tubes. They showed that these waves are the ideal magnetoseismological tool for probing radial plasma inhomogeneity in solar waveguides.
\citet{Karami2011} studied the effects of both density stratification and magnetic field expansion on torsional Alfv\'{e}n
waves in coronal loops. They concluded that the density stratification and magnetic field expansion have
opposite effects on the oscillating properties of torsional Alfv\'{e}n waves.

In the present work, we study the torsional Alfv\'{e}n waves, and determine the period ratio $P_{1}/P_{2}$ and
eigenfunctions in solar spicules.

\section{Theoretical modeling}
\label{sec:theory}
We consider an equilibrium configuration in the form of an expanding straight magnetic flux tube with varying density along tube.
We use cylindrical coordinates $r$, $\varphi$, and $z$ with the $z$-axis coinciding along tube axis. \citet{Okamoto2011}
claim that about $20\%$ of observed spicule kink waves are standing. Possibly there is a similar percentage of standing
torsional Alfv\'{e}n waves in spicules. So, in what follows we continue on standing torsional Alfv\'{e}n waves with the
nodes located at $z=0$ and $z=L$ ($L$ is spicule length).

To describe the plasma motion we use the linear ideal MHD equations for a cold plasma,

\begin{equation}
\label{eq:velo}
 \frac{\partial^{2} \mathbf{\xi}}{\partial t^{2}} = \frac{1}{\mu_{0}\rho} (\nabla \times \mathbf{b})\times
\mathbf{B},
\end{equation}
and

\begin{equation}
\label{eq:mag}
\mathbf{ b} = \nabla \times(\mathbf{\xi} \times \mathbf{B}),
\end{equation}
where $\mathbf{\xi}=(0,\xi_{\varphi},0)$ is the plasma displacement, and $\mathbf{b}=(0,b_{\varphi},0)$
is the magnetic field perturbation. Rewriting Eqs~\ref{eq:velo},~\ref{eq:mag} in components
and considering the time-dependence as $e^{-i\omega t}$ yields:
\begin{equation}
\label{eq:velo1}
 \mu_{0} \rho \omega^{2} \xi_{\varphi} + \frac{B_{r}}{r} \frac{\partial (rb_{\varphi})}{\partial r} + B_{z} \frac{\partial b_{\varphi}}{\partial z} = 0,
\end{equation}
and
\begin{equation}
\label{eq:mag1}
b_{\varphi} = \frac{\partial (B_{r}\xi_{\varphi})}{\partial r} + \frac{\partial (B_{z}\xi_{\varphi})}{\partial z}.
\end{equation}

Now like \citet{Verth2010,Ruderman2008,Karami2011} we use non-orthogonal flux coordinate in which $\psi$ becomes
an independent variable instead of $r$, i.e. $r=r(\psi,z)$. For an arbitrary function $f$ we have the following relations:
\begin{eqnarray}
\label{eq:f1}
  B_{r} &=& rB_{z} \frac{\partial f}{\partial \psi} \nonumber\\
  B_{z} &=& \frac{\partial f}{\partial z} - rB_{r} \frac{\partial f}{\partial \psi}.
\end{eqnarray}
When deriving these equations we assumed thin tube approximation (or long-wavelength limit), i.e. $R/L\ll 1$ ($R$ is the radius of expanding spicule).
The equilibrium magnetic field has two components, $r$, $z$,
and independent of $\varphi$, so that $B=B(r,z)$. It follows from solenoidal condition that $B$ can be expressed in terms of flux
function $\psi$,

\begin{eqnarray}
\label{eq:f2}
  \frac{\partial f}{\partial r} &=& -\frac{1}{r} \frac{\partial \psi}{\partial z} \nonumber\\
  \frac{\partial f}{\partial z} &=& \frac{1}{r} \frac{\partial \psi}{\partial r}.
\end{eqnarray}
Differentiating the identities $\psi=\psi(r(\psi,z),z)$ and $r=r(\psi(r,z),z)$ with respect to $z$ and using equation~\ref{eq:f2} yields:
\begin{eqnarray}
\label{eq:f3}
  \frac{\partial r}{\partial z} &=& \frac{B_{r}}{B_{z}} \nonumber\\
  \frac{\partial r}{\partial \psi} &=& \frac{1}{rB_{z}}.
\end{eqnarray}

By using Eqs~\ref{eq:f3}, along with the solenoidal condition,
we can simplify Eqs~\ref{eq:velo1},~\ref{eq:mag1} to

\begin{equation}
\label{eq:velo2}
 \mu_{0} \rho \omega^{2} \xi_{\varphi} + \frac{B_{z}}{R} \frac{\partial (R b_{\varphi})}{\partial z}= 0,
\end{equation}

and

\begin{equation}
\label{eq:mag2}
b_{\varphi} = R B_{z} \frac{\partial}{\partial z} \left(\frac{\xi_{\varphi}}{R}\right).
\end{equation}

Eqs~\ref{eq:velo2},~\ref{eq:mag2} can now be combined as

\begin{equation}
\label{eq:full}
 \mu_{0} \rho \omega^{2} \xi_{\varphi} + \frac{B_{z}}{R}\frac{\partial}{\partial z}
 \left[R^{2} B_{z} \frac{\partial}{\partial z} \left(\frac{\xi_{\varphi}}{R}\right)\right] = 0.
\end{equation}

It would be more realistic to use the background magnetic field, plasma density, and spicule radius inferred from the actual
magnetoseismology of observation \citep{Verth2011}:

\begin{eqnarray}
\label{eq:equlib}
  B_{z}(z) &=& B_{0}\exp (-z/H_{B}) \nonumber\\
  \rho(z) &=& \rho_{0}\exp (-z/H_{\rho}) \nonumber\\
  R(z) &=& R_{0}\exp (z/2H_{B}),
\end{eqnarray}
where $H_{B}$ and $H_{\rho}$ are magnetic and density scale heights, respectively. The latter follows from the conservation of magnetic
flux. Using Eqs~\ref{eq:equlib} in equation~\ref{eq:full} results in

\begin{equation}
\label{eq:pde}
 \frac{\partial^{2} \xi_{\varphi}}{\partial z^{2}} - \frac{1}{H_{B}}\frac{\partial \xi_{\varphi}}{\partial z} +
 \left( \frac{1}{4H^{2}_{B}} + 4 \pi ^{2} \omega ^{2} e^{-\alpha z}\right)\xi_{\varphi} =0,
\end{equation}
where $\alpha\equiv\left(\frac{H_{B} - 2H_{\rho}}{H_{\rho} H_{B}}\right)$.
In this equation the lengths are normalized to spicule length ($L$), and frequencies to Alfv\'{e}n frequency
($\omega_{A}\equiv\frac{V_{A}}{L}=0.06$ rad/s; $V_{A}=\frac{B_{0}}{\sqrt{\mu_{0} \rho_{0}}}=75$ km/s; $B_{0}=12$ G,
$\rho_{0}=1.9\times10^{-10}$ kg m$^{-3}$, $\mu_{0}=4\pi\times10^{-7}$ T m A$^{-1}$, and $L=8000$ km).
Magnetic and density scale heights are determined as $H_{B}=1816$ km and $H_{\rho}=752$ km by \citet{Verth2011}.

\section{Numerical results and Discussions}
\label{sec:resdis}
\begin{figure}
\centering
\includegraphics[width=8cm]{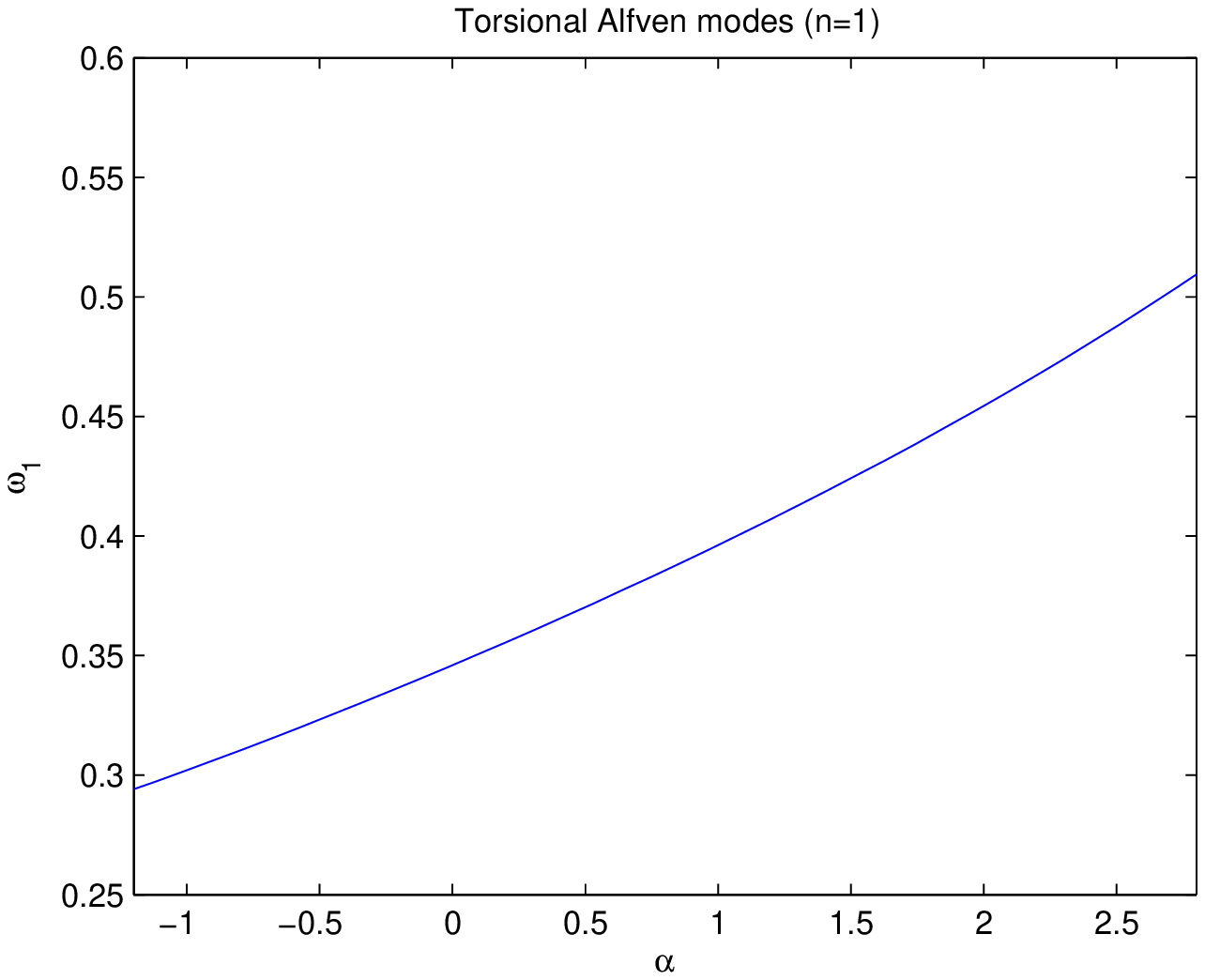}
\includegraphics[width=8cm]{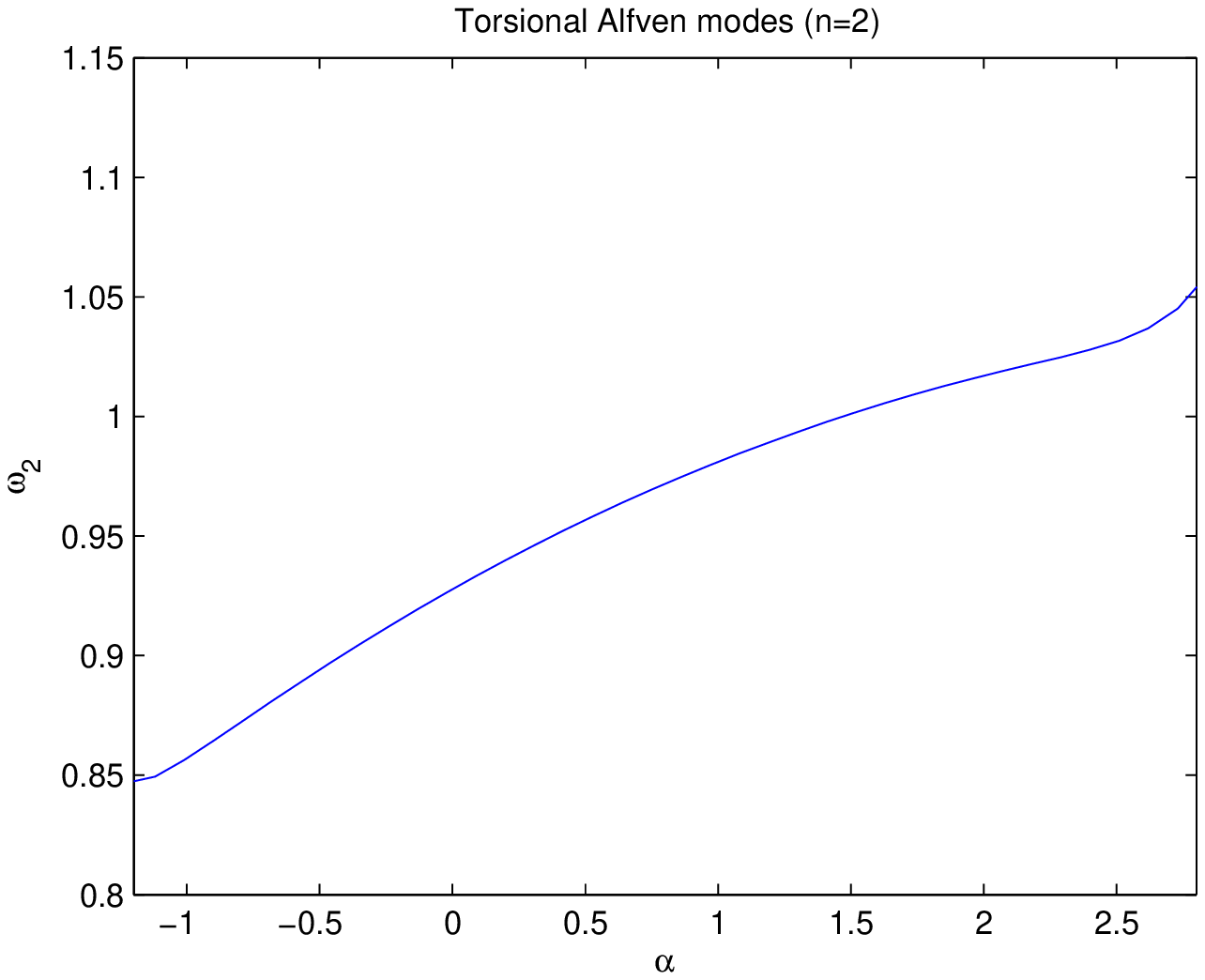}
\includegraphics[width=8cm]{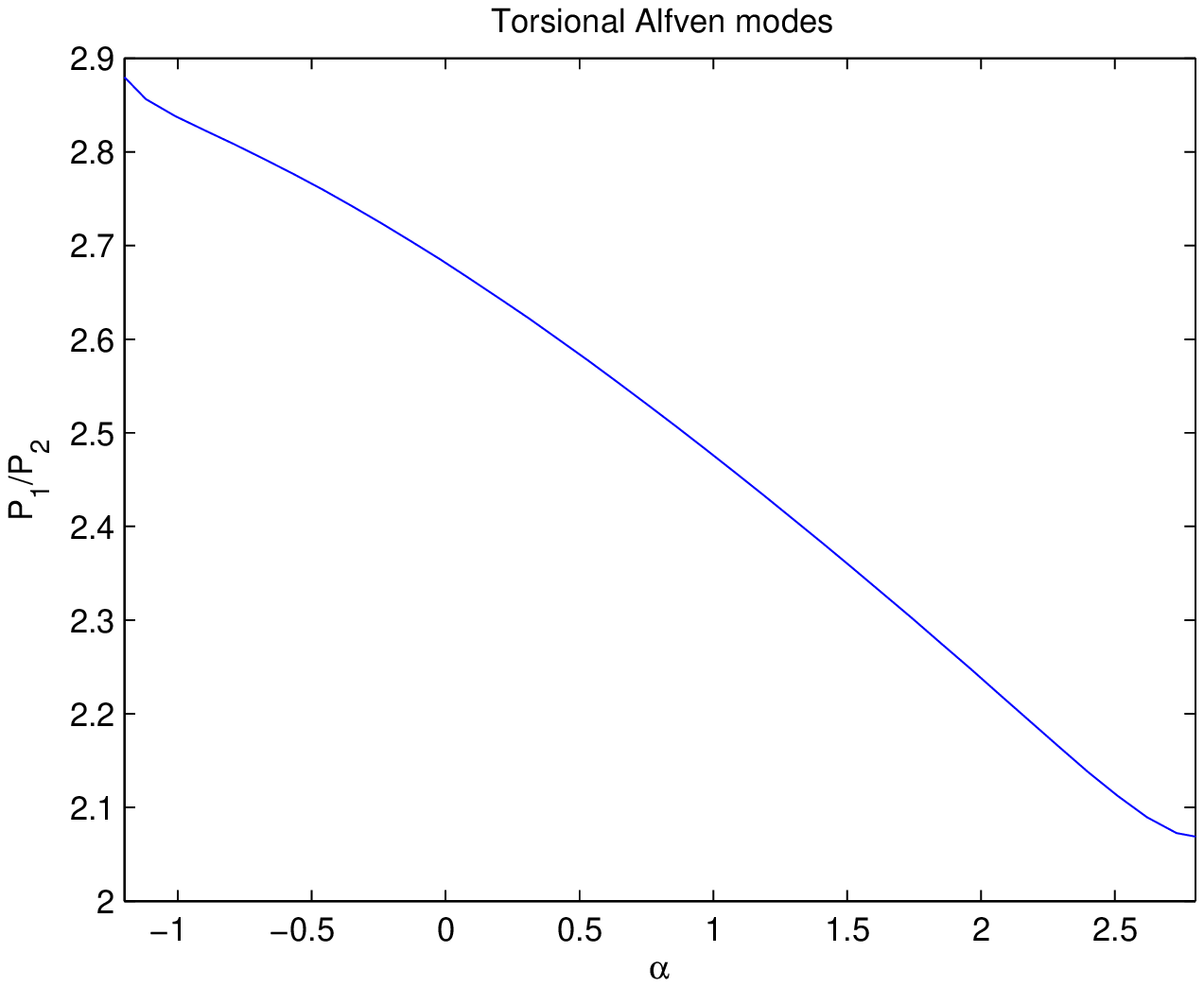}
\caption{Torsional Alfv\'{e}n modes frequencies and the period ratio $P_{1}/P_{2}$ between the period
$P_{1}$ of the fundamental mode and the period $P_{2}$ of its first harmonic are plotted from top to down panels, respectively. \label{fig1}}
\end{figure}

We solve equation~\ref{eq:pde} numerically by using differential transform method (DTM)(see appendix for detailed description of this method) to obtain both eigenfrequencies and eigenfunctions of standing torsional Alfv\'{e}n
waves in stratified and expanding solar spicules. We use the rigid boundary conditions and assume that $\xi_{\varphi}(0)=\xi_{\varphi}(L)=0$.
In Figure~\ref{fig1} we plotted Torsional Alfv\'{e}n modes frequencies and the period ratio $P_{1}/P_{2}$ between the period
$P_{1}$ of the fundamental mode and the period $P_{2}$ of its first harmonic. Frequencies are increasing with an increase of $\alpha$.
The ratio $P_{1}/P_{2}$ is decreasing with $\alpha$ and reaching to observed values around $\alpha=3$ $(H_{B} \simeq 3H_{\rho})$.
The fundamental mode and its first harmonic period ratios have departures from
its canonical value of $2$. It is distinguished by observations which were made by
\citet{Ebadi2014} in solar spicules. \citet{Andries2009} showed that the frequencies of detected kink
oscillation overtones are in particular sensitive to the density and the magnetic field expansion.
\citet{Karami2009} showed that the frequencies
and the damping rates of both the kink and fluting modes increase when the twist parameter
increases. These two factors are studied in spicules both observationally and
theoretically. \citet{Verth2011} based on SOT/\emph{Hinode} observations determined the variation of
magnetic field strength and plasma density along a spicule by seismology. They studied a kink wave
propagating along a spicule by estimating the spatial change in phase speed and velocity amplitude
as a novel approach. \citet{Suematsu2008,Tavabi2011} by using the SOT/\emph{Hinode} observations
reported twisted motions in spicules.

\begin{figure}
\centering
\includegraphics[width=8cm]{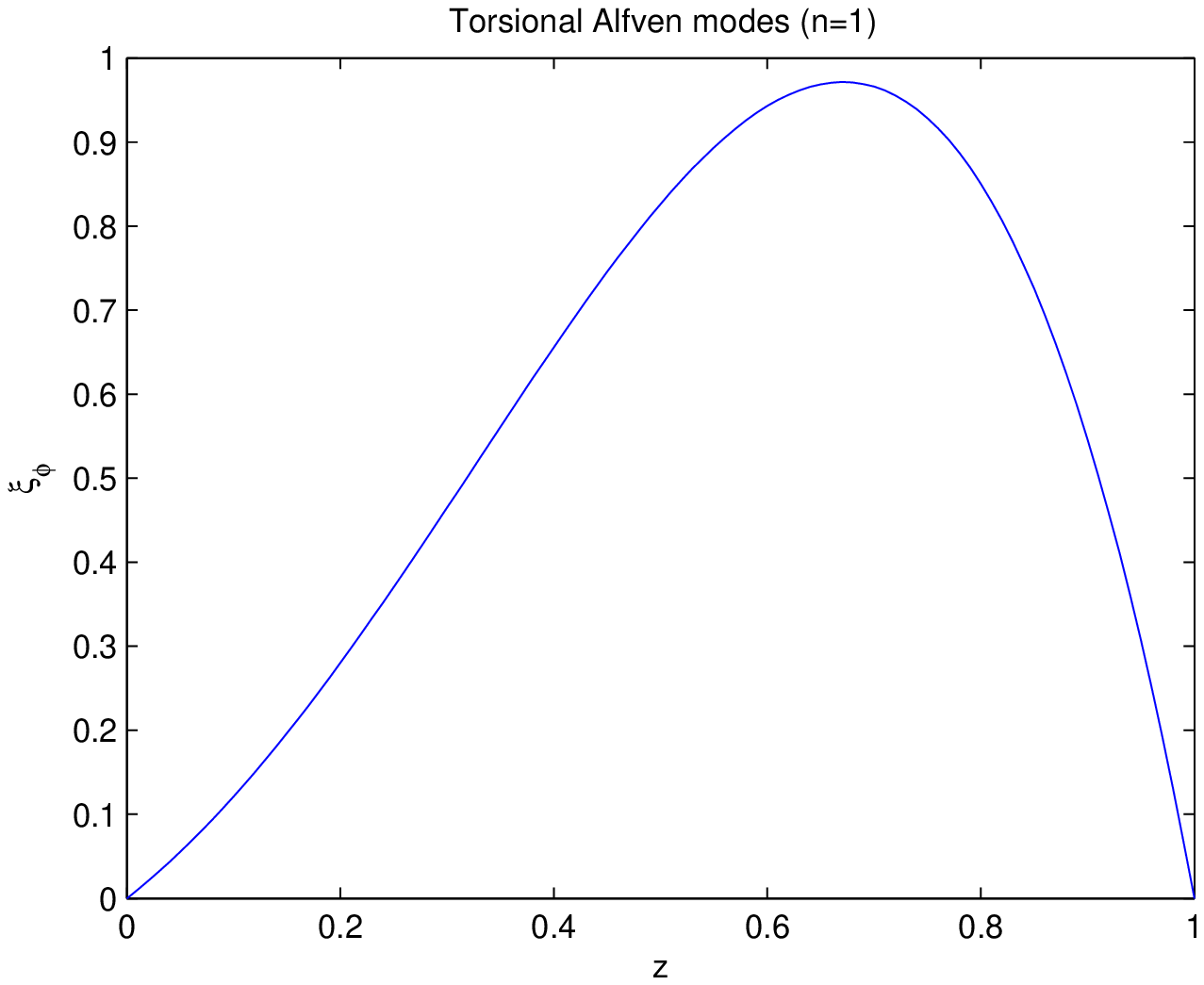}
\includegraphics[width=8cm]{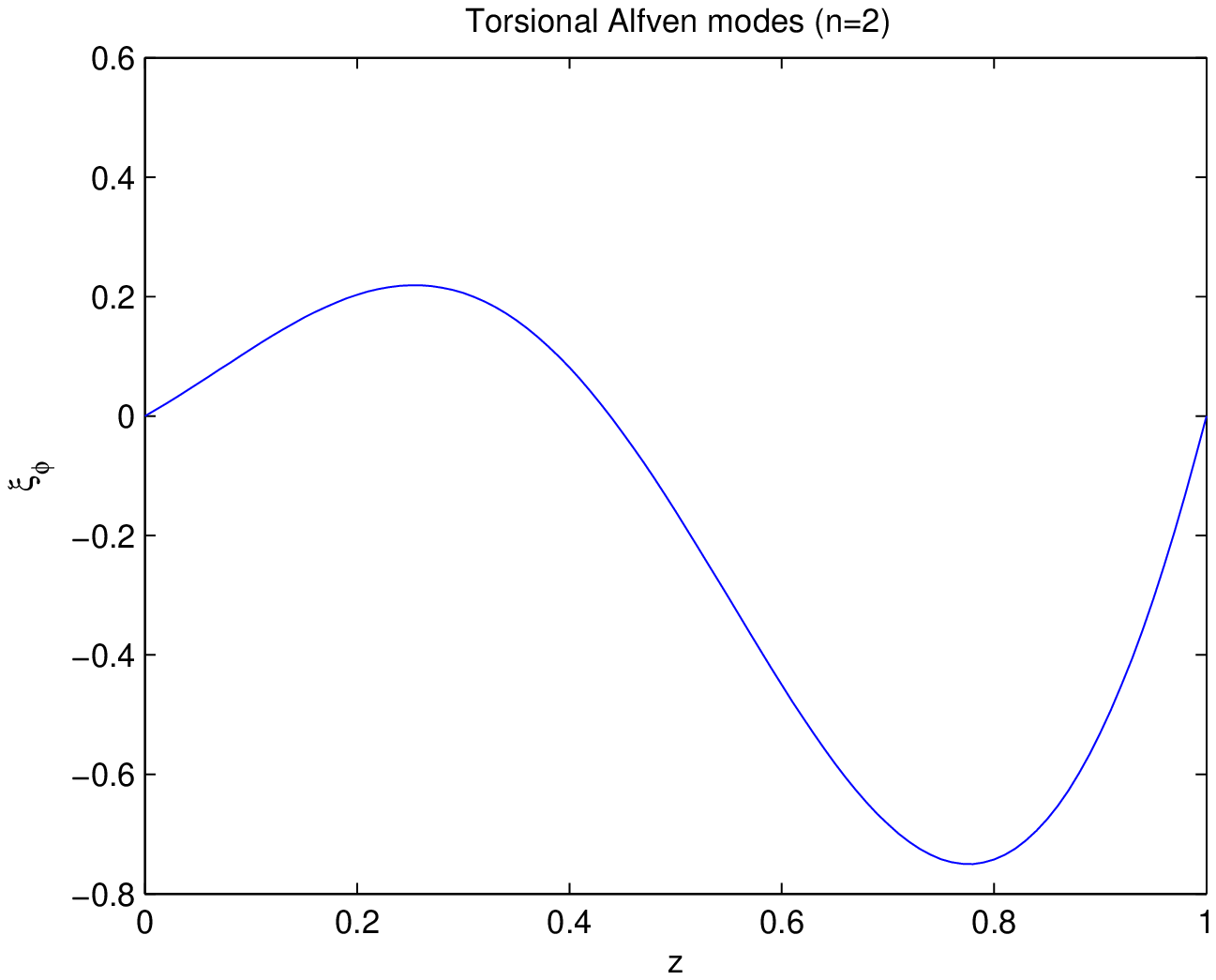}
\caption{Eigenfunctions of the fundamental and first harmonic torsional Alfv\'{e}n modes with respect to normalized height along the spicule
are plotted in the first and second panels, respectively ($\alpha=1.84$). \label{fig2}}
\end{figure}

Eigenfunctions of the fundamental and first harmonic torsional Alfv\'{e}n modes with respect to the normalized height along the spicule
are presented in Figure~\ref{fig2}. It should be emphasized that we plotted eigenfunctions for $\alpha=1.84$ which is determined by
observations. It is interesting that the oscillations amplitude are increasing towards higher heights. The same behavior was observed in spicules
by \citet{he2009,Ebadi2012a,Ebadi2014}. This means that with a little increase in height,
amplitude of oscillations become expanded due to significant decrease
in density, which acts as inertia against oscillations.

\section{Conclusion}
\label{sec:concl}
We consider an equilibrium configuration in the form of an expanding straight magnetic flux tube with varying density along tube.
We use cylindrical coordinates $r$, $\varphi$, and $z$ with the $z$-axis coinciding along tube axis. It is claimed that
about $20\%$ of observed spicule waves, are standing torsional Alfv\'{e}n waves. More realistic background magnetic field,
plasma density, and spicule radios inferred from the actual magnetoseismology of observations are used.
We used a novel mathematical method which was explained in the last section to solve equation~\ref{eq:pde}.
Fundamental and its first harmonic frequencies are increasing with $\alpha$. On the other hand their ratio is decreasing with $\alpha$
and reaching to the observed values around $\alpha \simeq 3$. The fundamental mode and its first harmonic period ratios have departures from
its canonical value of $2$ which was distinguished by observations. The density stratification and magnetic twist are two main factors which
make the period ratio departures from its canonical value of $2$. These two factors are studied in spicules both observationally and
theoretically. Eigenfunction variations with height show that the oscillations amplitude are increasing towards higher heights. It is in agreement
with the results of spicule observations. This means that with a little increase in height, amplitude of oscillations become expanded due to
significant decrease in density, which acts as inertia against oscillations.

\acknowledgments
This work has been supported financially by the Research Institute for Astronomy and
Astrophysics of Maragha (RIAAM), Maragha, Iran.

\section{Appendix}
\subsection{Differential Transform Method}
The differential transform technique is one of the semi-numerical analytical methods for ordinary and partial differential equations that uses the polynomials as approximations of the exact solutions that are sufficiently differentiable \citep{Erturk2012}. The basic definition and fundamental theorems of the differential
transform method (DTM) and its applicability for various kinds of differential and integral equations are given in \cite{Nazari2010, Tari2011, Ozkan2010,Arikogli2006}. For convenience of the reader,
we present a review of the DTM. The differential transform of the $k$th derivative of function $f(t)$ is defined as follows:
\begin{equation}\label{A1}
F(k)=\frac{1}{k!}\left[ \frac{d^kf(t)}{dt^k} \right]_{t=t0},
\end{equation}
where $f(t)$ is the original function and $F(k)$ is the transformed function. The differential inverse transform
of $F(k)$ is defined as
\begin{equation}\label{A2}
f(t)=\sum_{k=0}^\infty F(k)(t-t_0)^k.
\end{equation}
From Eqs. \ref{A1} and \ref{A2}, we get
\begin{equation}\label{A3}
f(t)=\sum_{k=0}^\infty \frac{(t-t_0)^k}{k!}\frac{d^kf(t)}{dt^k} \|_{t=t0},
\end{equation}
which implies that the concept of differential transform is derived from Taylor series expansion,
but the method dose not evaluate the derivative symbolically. However, relative derivatives are calculated by an iterative
way which are described by the transformed equations of the original function. For implementation purpose, the function $f(t)$
is expressed  by a finite series and Equation~\ref{A2} can be written as
\begin{equation}\label{A4}
f(t)=\sum_{k=0}^N F(k)(t-t_0)^k.
\end{equation}
Here $N$ is decided by the convergence of natural frequency. The  fundamental operations performed by differential transform can
readily be obtained and are listed in Table 1. The main steps of the DTM, as a tool for solving different classes of nonlinear problems,
are the following. First, we apply the differential transform (Equation~\ref{A1}) to the given problem (integral equation, ordinary differential
equation or partial differential equation), then the result is a recurrence relation. Second, solving this relation and using the differential
inverse transform (Equation~\ref{A4}), we can obtain approximate solution of the problem.

Table 1: Operations of differential transform
$$
\tiny{
\begin{array}{cc}
 \hline
 \text{Original function}  & \text{Transformed function} \\ \hline
f(t)=u(t)\pm v(t)  & F(k)=U(k)\pm V(k) \\ \hline
 f(t)=u(t)v(t)  & F(k)=\sum_{l=0}^k U(l)\times V(k-l)   \\ \hline
 f(t)=\alpha u(t)  & F(k)=\alpha U(k) \\ \hline
 f(t)=\frac{du(t)}{dt}    & F(k)=(k+1)U(k+1) \\ \hline
f(t)=\frac{d^mu(t)}{{dt}^m}   & F(k)=(k+1)(k+2)\cdots (k+m)U(k+m)  \\ \hline
f(t)=\int_{t_0}^t u(t)dt  & F(k)=\frac{U(k-1)}{k}, \ \ k\geq 1  \\ \hline
f(t)=t^m & F(k)=\delta(k-m)=\left\{
\begin{array}{c}
 1, \ \ \  k=m,\\
 0, \ \ \ k\neq m.
\end{array}
\right.  \\ \hline
   f(t)=exp(\lambda,t)  & F(k)=\frac{\lambda^k}{k!}  \\ \hline
    f(t)=sin(\omega t+\alpha)   & F(k)=\frac{\omega^k}{k!}sin(\frac{\pi k}{2}+alpha) \\ \hline
     f(t)=cos(\omega t+\alpha)   & F(k)=\frac{\omega^k}{k!}cos(\frac{\pi k}{2}+alpha)   \\ \hline
   f(t)=\frac{u(t)}{v(t)}  & F(k)=\frac{1}{V(0)}\left[U(k)-\sum_{m=0}^{k-1}U(k)V(k-m)\right]  \\ \hline
    f(t)={\left[u(t)\right]}^b  & F(k)=\left\{
    \begin{array}{c}
    U(0),\ \ \  k=0,\\
    \sum_{m=0}^k\frac{(b+1)m-k}{k U(0)}U(m)U(k-m), \ \ \ k\geq 1.
    \end{array}
    \right.\\
          \hline
\end{array}}$$

\makeatletter
\let\clear@thebibliography@page=\relax
\makeatother

\end{document}